\documentclass[aps,preprint]{revtex4}
\usepackage{amssymb,epsf}
\usepackage{amsmath}
\usepackage{graphicx}
\begin{document}
\title{Topological Black Holes of Einstein-Yang-Mills dilaton Gravity}
\author{M. H. Dehghani$^{1,2}$\footnote{email address:
mhd@shirazu.ac.ir} and A. Bazrafshan$^1$}
\affiliation{$^1$ Physics Department and Biruni Observatory, College of Sciences, Shiraz University, Shiraz 71454, Iran\\
$^2$ Research Institute for Astrophysics and Astronomy of Maragha (RIAAM), Maragha, Iran}

\begin{abstract}
We present the topological solutions of Einstein-dilaton gravity in the presence
of a non-Abelian Yang-Mills field. In $4$ dimensions, we consider the
$So(3)$ and $So(2,1)$ semisimple group as the Yang-Mills gauge group, and
introduce the black hole solutions with spherical and hyperbolic horizons, respectively.
The solution in the absence of dilaton potential is asymptotically flat
and exists only with spherical horizon.
Contrary to the non-extreme Reissner-Nordstrom
black hole, which has two horizons with a timelike and avoidable singularity,
here the solution may present a black hole with a null and
unavoidable singularity with only one horizon. In the
presence of dilaton potential, the asymptotic behavior of the solutions
is neither flat nor anti-de Sitter. These solutions contain a null and
avoidable singularity, and may present a black hole with two
horizons, an extreme black hole or a naked singularity. We also calculate the mass of the solutions
through the use of a modified version of Brown and York formalism, and consider the
first law of thermodynamics.

\end{abstract}

\maketitle

\section{Introduction}

In the low energy limit of the string theory, one recovers Einstein gravity
along with a scalar dilaton field which is non-minimally coupled to a matter
field \cite{Wit1}. A typical feature of the dilaton in string theory is its
exponential coupling to the matter field in the frame where the dilaton
decouples from the Ricci scaler. Exact solutions for charged dilaton
black holes in which the dilaton is coupled to the Maxwell field have been
constructed by many authors. It is found that the presence of dilaton has
important consequences on the asymptotic behavior and the thermodynamic
properties of the black hole solutions. The asymptotically flat solutions of
Einstein-Maxwell-dilaton (EMD) theory have been investigated in \cite{CD}.
Solutions of EMD theory with one Liouville-type potential which are neither
asymptotically flat nor anti-de Sitter (AdS) have been considered in \cite
{AdSdil}. These kinds of solutions with three Liouville-type potentials
have been considered in \cite{Gao}.

Although the gauged supergravity AdS
theories generically contain Yang-Mills (YM) fields, most of the studies in
the literature have been restricted to the case of Abelian gauge fields in the
bulk. In this work we consider Einstein-dilaton gravity in the presence of the
non-Abelian Yang-Mills field and investigate the existence of exact
solutions with different asymptotic behavior. In the absence of dilaton,
this theory with $SU(2)$ gauge group has both solitonic \cite{BK} and
colored black hole solutions \cite{YMblack}, while Einstein-Yang-Mills (EYM)
with $So(N)$ or $So(N-1,1)$\ gauge groups has only black hole solutions \cite
{Yas,Hal1,BD}. These solutions have led to certain revisions of some of the
basic concepts of black hole physics based on the uniqueness and no-hair
theorem. It is now well-known that this theory possesses ''hairy'' black
hole solutions, whose metric is not a member of the Kerr-Newmann family (see
\cite{Volkov} for a detailed review in 4 dimensions and \cite{Radu1} for a
recent review in higher dimensions). Solutions of the EYM equations were
also investigated in the presence of cosmological constant \cite{Vol2,Eliza}%
. In the presence of dilaton, the solutions of Einstein-Yang-Mills-dilaton
(EYMD) theory have been considered by many authors \cite{EYMd}. The presence
of a dilaton field also invites the topological black holes into the game.
Indeed, the horizon topology of an asymptotically flat black hole should be
a round sphere, while in the presence of dilaton with a potential, it is
possible to have a black holes with zero or negative constant curvature
horizon too. These black holes are referred to as topological black holes in
the literature. These kinds of solutions in EMD theory have been investigated
in \cite{Cai}. Here, we present the black hole solutions of Einstein-dilaton
gravity in the presence of non-Abelian Yang-Mills field with $SO(3)$ and $%
SO(2,1)$ gauge group with different asymptotic behavior. We use the
Wu-Yang ansatz \cite{Wu} in which the Yang-Mills (YM) gauge
potential have only angular components.

The outline of the work is as follows. In Sec. \ref
{Fiel}, we give a brief review of the field
equations of Einstein-dilaton gravity in the presence of YM gauge fields
with a semisimple gauge group. In Sec. \ref{Sph} we present static solutions
with spherical and hyperbolic horizons for gauge fields with $So(3)$ and $So(2,1)$ gauge group.
We calculate the mass of the solutions through the use of a modified
Brown and York formalism and investigate the first law of thermodynamics. Finally, we give some concluding remarks.

\section{Field equations\label{Fiel}}

Here, we review the field equations of Einstein gravity coupled with dilaton
and non-Abelian Yang-Mills field with an $N$-parameters gauge group ${\cal G}
$, which is assumed to be at least semisimple with structure constants $%
C_{bc}^{a}$ and metric tensor $\Gamma _{ab}=C_{ad}^{c}C_{bc}^{d}$. The
action of this theory may be written as
\begin{eqnarray}
I_{G} &=&-\frac{1}{16\pi }\int_{{\cal M}}d^{n+1}x\sqrt{-g}\left( {\cal R}%
\text{ }-\frac{4}{n-1}\partial _{\mu }\Phi \partial ^{\mu }\Phi -V(\Phi
)-e^{-4\alpha \Phi /(n-1)}\gamma _{ab}F_{\mu \nu }^{(a)}F^{(b)\mu \nu
}\right)  \nonumber \\
&&-\frac{1}{8\pi }\int_{\partial {\cal M}}d^{n}x\sqrt{-h}K,  \label{Act}
\end{eqnarray}
where the Latin indices $a$, $b$.... go from $1$ to $N$, and the repeated
indices is understood to be summed over, ${\cal R}$ is the Ricci scalar
curvature, $\Phi $ is the dilaton field, $V(\Phi )$ is a potential for $\Phi
$ and
\begin{equation}
\gamma _{ab}\equiv -\frac{\Gamma _{ab}}{\left| \det \Gamma _{ab}\right|
^{1/N}}.  \label{gam}
\end{equation}
In Eq. (\ref{Act}) $F_{\mu \nu }^{(a)}$'s are the non-Abelian gauge fields:
\begin{equation}
F_{\mu \nu }^{\left( a\right) }=\partial _{\mu }A_{\nu }^{\left( a\right)
}-\partial _{\nu }A_{\mu }^{\left( a\right) }+\frac{1}{2e}C_{bc}^{a}A_{\mu
}^{\left( b\right) }A_{\nu }^{\left( c\right) },  \label{gfield}
\end{equation}
where $e$ is a coupling constant and $A_{\mu }^{\left( a\right)
}$'s are the gauge potentials.
The last term in Eq. (\ref{Act}) is the
Gibbons-Hawking boundary term which is chosen such that the variational
principle is well-defined, and $K$ is the trace of the extrinsic curvature $%
K^{ab}$ of any boundary(ies) $\partial {\cal M}$ of the manifold ${\cal M}$,
with induced metric(s) $h_{ab}$. Variation of the action (\ref{Act}) with
respect to the gauge potential $A_{\mu
}^{(a)}$, the spacetime metric $g_{\mu \nu }$ and the dilaton field $\Phi $ yield the EYMD equations as
\begin{eqnarray}
F^{(a)\mu \nu }_{\text{\ \ \ \ \ };\nu} &=&\frac{4\alpha}{n-1} F^{(a)\mu \nu}\Phi_{;\nu}
+\frac{1}{e}
C_{bc}^{a}A_{\nu }^{(b)}F^{(c)\nu \mu },  \label{YMeq} \\
{\cal R}_{\mu \nu }&=&\frac{4}{n-1}\left( \partial _{\mu }\Phi \partial
_{\nu }\Phi +\frac{1}{4}g_{\mu \nu }V(\Phi )\right) \nonumber\\
&& +2\exp \left( -\frac{%
4\alpha \Phi }{n-1}\right) \gamma _{ab}\left( F_{\mu }^{(a)\lambda }F_{\nu
\lambda }^{(b)}-\frac{1}{2(n-1)}g_{\mu \nu }F^{(a)\lambda \sigma }F_{\lambda
\sigma }^{(b)}\right) ,  \label{EDeq} \\
\nabla ^{2}\Phi &=&\frac{n-1}{8}\frac{\partial V}{\partial \Phi }-\frac{%
\alpha }{2}\exp \left( -\frac{4\alpha \Phi }{n-1}\right) \gamma
_{ab}F_{\lambda \eta }^{(a)}F^{(b)\lambda \eta },  \label{Dileq}
\end{eqnarray}

\section{Static Black Holes\label{Sph}}

The 4-dimensional metric of a spherically symmetric spacetime may be written as
\begin{equation}
ds^{2}=-f(r)dt^{2}+\frac{dr^{2}}{f(r)}+R^{2}(r)d\Omega ^{2},
\label{Met1}
\end{equation}
where
\begin{eqnarray}
d\Omega ^{2} &=&d\theta ^{2}+\sin ^{2}\theta d\varphi^2 ;\text{ \ \ \ \ \ }k=1,
\nonumber \\
&=&d\theta ^{2}+\sinh ^{2}\theta d\varphi^2 ;\text{ \ \ \ }k=-1,  \label{MetB}
\end{eqnarray}
for spherical and hyperbolic horizons, respectively. Here, we consider the
static case, and therefore we assume that $\Phi $ depends on $r$ only. For
the spacetime with spherical horizon ($k=1$), the\ $So(3)$ YM potentials
\begin{eqnarray}
A_{\mu }^{(1)} &=&e\left( -\cos \varphi d\theta +\sin \theta \cos \theta
\sin \varphi d\varphi \right) ,  \nonumber \\
A_{\mu }^{(2)} &=&-e\left( \sin \varphi d\theta +\sin \theta \cos \theta
\cos \varphi d\varphi \right) ,  \nonumber \\
A_{\mu }^{(3)} &=&e\sin ^{2}\theta d\varphi ,  \label{Asph}
\end{eqnarray}
with non-zero components of YM field tensor as
\begin{eqnarray}
F^{(1)\theta \varphi } &=&-\frac{e\sin \varphi }{R^{4}(r)},  \nonumber
\\
F^{(2)\theta \varphi } &=&\frac{e\cos \varphi }{R^{4}(r)},  \nonumber \\
F^{(3)\theta \varphi } &=&\frac{e\cot \theta }{R^{4}(r)}.  \label{Ften}
\end{eqnarray}
satisfy the YM equation (\ref{YMeq}). For the spacetime with hyperbolic
horizon, $k=-1$, the $So(2,1)$ gauge fields
\begin{eqnarray}
A_{\mu }^{(1)} &=&e\left( -\cos \varphi d\theta +\sinh \theta \cosh \theta
\sin \varphi d\varphi \right) ,  \nonumber \\
A_{\mu }^{(2)} &=&-e\left( \sin \varphi d\theta +\sinh \theta \cosh \theta
\cos \varphi d\varphi \right) ,  \nonumber \\
A_{\mu }^{(3)} &=&e\sinh ^{2}\theta d\varphi ,  \label{Ahyp}
\end{eqnarray}
satisfy the YM field equation (\ref{YMeq}).

In order to solve the field equation (\ref{EDeq}) for three unknown functions
$f(r)$, $R(r)$ and $\Phi (r)$, we make the ansatz
\begin{equation}
R(r)=r\exp \left( -\alpha \Phi \right) .  \label{Rphi}
\end{equation}
Using the above ansatz (\ref{Rphi}), the YM fields (\ref{Asph}) or (\ref{Ahyp})
and the metric (\ref{Met1}), one can easily show that the components
of Eq. (\ref{EDeq}) reduce to
\begin{eqnarray}
&& (r^2 f^{\prime})^{\prime }-2\alpha r^2 \Phi ^{\prime }f^{\prime
}+V r^2-2e^{2}r^{-2}\exp (2\alpha \Phi ) =0,  \label{Eq1} \\
&& (r^2 f^{\prime})^{\prime }-2\alpha r^2 \Phi ^{\prime }f^{\prime
}+V r^2-2e^{2}r^{-2}\exp (2\alpha \Phi )=4 r f \left[\alpha r \Phi ^{\prime \prime
}-2\alpha \Phi ^{\prime }+(1+\alpha ^{2})r \Phi ^{\prime ^{2}}\right],
\label{Eq2} \\
&& 2(rf)^{\prime }-2\alpha r^2 (f \Phi ^{\prime })^{\prime }+V r^2
+2(e^{2}r^{-2}-k)\exp (2\alpha \Phi )
+4\alpha rf\Phi ^{\prime }(\alpha r\Phi ^{\prime}-2) =0,
\label{Eq3}
\end{eqnarray}
where prime denotes the derivative with respect to $r$. Subtracting Eq. (\ref
{Eq1}) from Eq. (\ref{Eq2}) gives
\begin{equation}
\alpha r\Phi ^{\prime \prime }-(1+\alpha ^{2})r{\Phi ^{\prime }}^{2}+2\alpha
\Phi ^{\prime }=0,  \label{EqPhir}
\end{equation}
with the solution
\begin{equation}
\Phi (r)=-\frac{\alpha }{(1+\alpha ^{2})}\ln (1-\frac{r_{0}}{r}),
\label{Phir}
\end{equation}
where $r_{0}$ is an arbitrary constant. Thus, the metric (\ref{Met1}) may be
written as
\begin{equation}
ds^{2}=-f(r)dt^{2}+\frac{dr^{2}}{f(r)}+r^{2}\left( 1-\frac{r_{0}}{r}\right)
^{2\alpha ^{2}/(1+\alpha ^{2})}d\Omega ^{2},  \label{Met2}
\end{equation}
which is physical only for $r\geq r_{0}$. Thus, one should restrict the
spacetime to the region $r\geq r_{0}$, by introducing a new radial
coordinate $\rho $\ as:
\begin{equation}
\rho ^{2}=r^{2}-r_{0}^{2}\Rightarrow dr^{2}=\frac{\rho ^{2}}{\rho
^{2}+r_{0}^{2}}d\rho ^{2}  \label{Tr}
\end{equation}
With this new coordinate, the above metric becomes:
\begin{equation}
ds^{2}=-f(\rho )dt^{2}+\frac{\rho ^{2}d\rho ^{2}}{(\rho
^{2}+r_{0}^{2})f(\rho )}+(\rho ^{2}+r_{0}^{2})\left( 1-\frac{r_{0}}{\sqrt{%
\rho ^{2}+r_{0}^{2}}}\right) ^{2\alpha ^{2}/(1+\alpha ^{2})}d\Omega ^{2},
\label{Met3}
\end{equation}
which is now physical for $0\leq \rho <\infty $. In the rest of the paper,
we work in the $r$-coordinate for simplicity.

\subsection{Asymptotically flat solutions:}

\begin{figure}
\centering {\includegraphics[width=7cm]{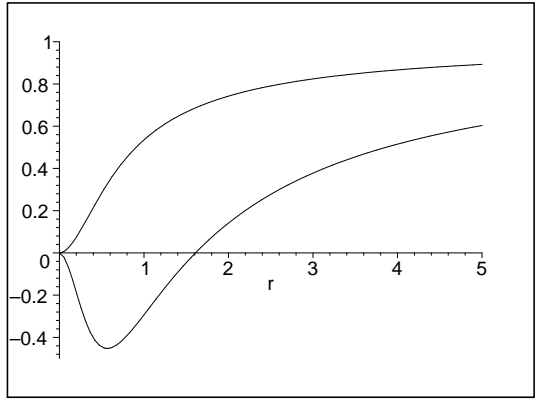} }
\caption{$f(r)$ versus $r$ for $\alpha=.2$, $r_0=.5$,
$e\leq e_{\mathrm{ext}}$ and $e>e_{\mathrm{ext}}$ from up
to down, respectively.} \label{Fig1}
\end{figure}

\begin{figure}
\centering {\includegraphics[width=7cm]{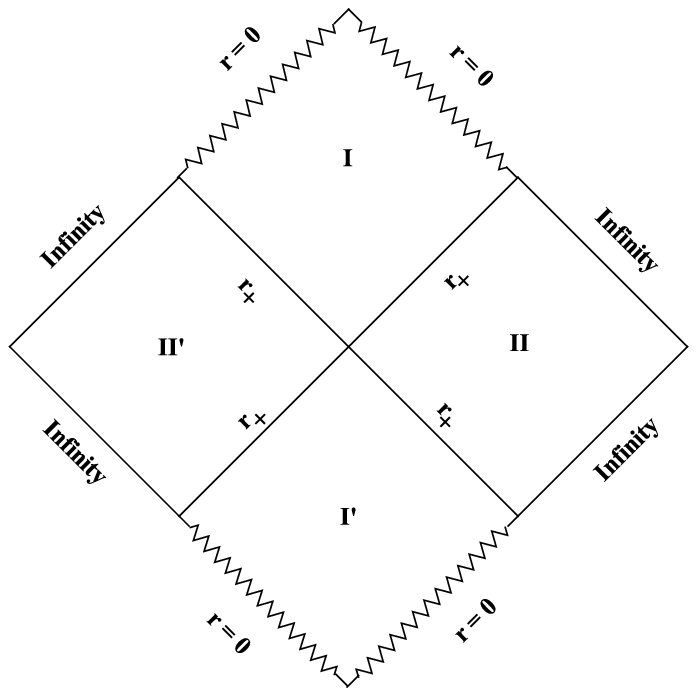} }
\caption{Penrose diagram for $e>e_{\mathrm{ext}}$.} \label{Pen1}
\end{figure}

First, we consider the solutions with $V(\Phi )=0$. For this case,
the solution exists only for $k=1$ with spherical horizon. Substituting $\Phi (r)$
in the field equations (\ref{Eq1})-(\ref{Eq3}) one finds the function $f(r)$
as
\begin{equation}
f(r)=\left(1-\frac{(1+\alpha ^{2})e^{2}}{r_{0}r}\right) \left( 1-\frac{r_{0}%
}{r}\right) ^{(1-\alpha ^{2})/(1+\alpha ^{2})}.  \label{fr1}
\end{equation}
One can show that the Kretschmann scalar $R_{\mu \nu \rho \sigma }R^{\mu \nu
\rho \sigma }$ diverges at $\rho =0$ $(r=r_{0})$, and therefore there is a
curvature singularity located at $r=r_{0}$. The solution presents a black
hole with horizon radius $r_{+}=(1+\alpha ^{2})e^{2}/r_{0}$, provided $%
e^{2}>r_{0}^{2}/(1+\alpha ^{2})$, and a naked singularity otherwise
(see Fig. \ref{Fig1}).
Contrary to the Reissner-Nordstrom solution, which is the $\alpha =0$ limit
of this solution with two horizons, the solution given by Eqs. (\ref{Met2}) and
(\ref{fr1}) has only one horizon. While the singularity of
Reissner-Nordstrom is timelike and avoidable, here the singularity is null and unavoidable.
This can be seen in the Penrose diagram of the solution, which is
drawn in Fig. \ref{Pen1}. The Hawking temperature of the
black holes can be easily obtained by requiring the absence of conical
singularity at the horizon in the Euclidean sector of the black hole
solutions. One obtains
\begin{equation}
{T}_{+}=\frac{1}{4\pi r_{+}}\left( 1-\frac{r_{0}}{r_{+}}\right) ^{(1-\alpha
^{2})/(1+\alpha ^{2})}.  \label{Temp1}
\end{equation}

\subsection{Asymptotically non-flat solutions:}

\begin{figure}
\centering {\includegraphics[width=7cm]{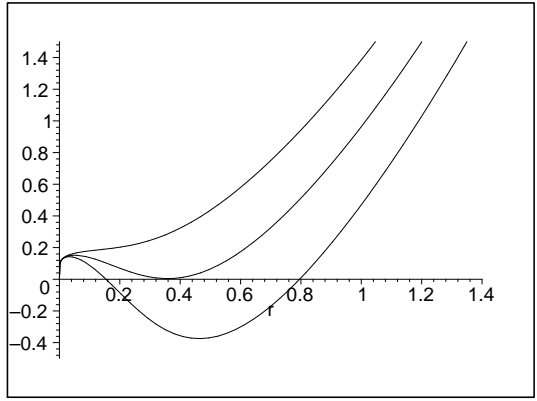} }
\caption{$f(r)$ versus $r$ for $\alpha=.2$, $r_0=.5$,
$e< e_{\mathrm{ext}}$, $e=e_{\mathrm{ext}}$ and $e>e_{\mathrm{ext}}$ from up
to down, respectively.} \label{Fig2}
\end{figure}

Here, we consider the solutions of EYMD theory in the presence of the
following potential \cite{Gao}
\begin{eqnarray}
V(\Phi )&=&-\frac{1}{(1+\alpha ^{2})l^{2}}\Bigg{\{}-2\alpha ^{2}(1-3\alpha ^{2})
exp\left(-\frac{2\Phi }{\alpha }\right)  \nonumber \\
&& +2(3-\alpha ^{2})\exp(2\alpha \Phi )+16\alpha ^{2}\exp\left(-\frac{\Phi
(1+\alpha ^{2})}{\alpha }\right)\Bigg{\}}.  \label{Vp}
\end{eqnarray}
Such a potential may arise from the compactification of a higher dimensional
supergravity model which originates from the low energy limit of
a background string theory \cite{Gid}. With this potential, the solution of Eqs. (\ref{Eq1})-(\ref{Eq3}) may be
written as
\begin{equation}
f(r)=\left( k-\frac{(1+\alpha ^{2})e^{2}}{r_{0}r}\right) \left( 1-\frac{r_{0}%
}{r}\right) ^{(1-\alpha ^{2})/(1+\alpha ^{2})}+\frac{r^{2}}{l^{2}}\left( 1-%
\frac{r_{0}}{r}\right) ^{2\alpha ^{2}/(1+\alpha ^{2})}.  \label{fr2}
\end{equation}
This metric at large $r$ may be written as:
\begin{equation}
ds^{2}=-f_0(r)dt^{2}+\frac{dr^{2}}{f_0(r)}+r^{2}d\Omega ^{2}, \label{dsAs}
\end{equation}
where
\begin{equation}
f_{0}(r)=k-\frac{\alpha^2 r_0^2}{l^2(1+\alpha^2)^2}+\left(\frac{r}{l}-\frac{\alpha^2}{1+\alpha^2}\frac{r_0}{l}\right)^2. \label{fr0}
\end{equation}
Thus, the asymptotic behavior of the spacetime in the presence of the
potential (\ref{Vp}) is not exactly AdS. Indeed the metric given by Eqs. (\ref{dsAs}) and (\ref{fr0})
does not satisfy the Einstein field equation in the presence of the cosmological constant.
Again the solution given by Eqs. (\ref{Met2}) and (\ref{fr2}) has a curvature singularity at $r=r_{0}$, and presents a
black hole with two horizons if $e>e_{{\rm ext}}$, an extreme black hole
when $e=e_{{\rm ext}}$, and a naked singularity otherwise, where
\begin{equation}
e_{{\rm ext}}^2=\frac{r_0 r_{{\rm ext}}}{1+\alpha ^{2}}\left\{ k+\frac{r_{{\rm ext}}^{2}}{l^{2}}%
\left( 1-\frac{r_{0}}{r_{\mathrm{ext}}}\right) ^{(3\alpha ^{2}-1)/(1+\alpha ^{2})}\right\}
. \label{qext}
\end{equation}
In Eq. (\ref{qext}) $r_{{\rm ext}}$ is the root of ${T}_{+}=0$, where ${T}%
_{+}$ is the temperature of the black hole give by
\begin{equation}
{T}_{+}=\frac{1}{4\pi r_{+}}\left( 1-\frac{r_{0}}{r_{+}}\right) ^{(1-\alpha
^{2})/(1+\alpha ^{2})}\left\{ k+\frac{3(1+\alpha ^{2})r_{+}^{2}-4br_{+}}{%
(1+\alpha ^{2})l^{2}}\left( 1-\frac{r_{0}}{r_{+}}\right) ^{-2(1-\alpha
^{2})/(1+\alpha ^{2})}\right\} .  \label{T2}
\end{equation}
Figure \ref{Fig2} shows the metric function as a function of $r$ for various values of $e$.
Here, again the singularity is null, but it is avoidable.

\section{Mass of the black holes}
It is well known that the gravitational action given in Eq. (\ref{Act})
diverges. A systematic method of dealing with this divergence for
asymptotically AdS solutions is through the use of the counterterm method
inspired by AdS/CFT correspondence. However, our solutions (\ref{fr1}) or (%
\ref{fr2}) are not asymptotically AdS. Thus, we use the substraction method
of Brown and York (BY) \cite{BY}. In order to use the BY method the metric should
have the form
\begin{equation}
ds^{2}=-W({R})dt^{2}+\frac{d{R}^{2}}{V(R)}+R%
^{2}d\Omega ^{2}. \label{BYmet}
\end{equation}
Thus, we should first write the metric (\ref{Met2}) in the above form. To do
this, we perform the following transformation:
\[
R=r\left( 1-\frac{r_{0}}{r}\right) ^{\alpha ^{2}/(1+\alpha ^{2})}.
\]
It is a matter of calculations to show that the metric (\ref{Met2}) gets the same
form as (\ref{BYmet}) with the following metric functions:
\begin{eqnarray}
&& W(R)=f(r(R)),\\
&& V({R)}=f(r(R))\left( \frac{dR}{dr}\right) ^{2}=\left( 1-%
\frac{r_{0}}{(1+\alpha ^{2})r}\right) \left( 1-\frac{r_{0}}{r}\right)
^{-1/(1+\alpha ^{2})}f(r(R)).
\end{eqnarray}
The background metric is chosen to be the metric (\ref{BYmet}) with:
\begin{eqnarray}
&& W_{0}(R)=k-\frac{\alpha^2 r_0^2}{l^2(1+\alpha^2)^2}+\left(\frac{R}{l}-\frac{\alpha^2}{1+\alpha^2}\frac{r_0}{l}\right)^2,\\
&& V_{0}(R)=k+\left(\frac{R}{l}-\frac{\alpha^2}{1+\alpha^2}\frac{r_0}{l}\right)^2. \label{Asym}
\end{eqnarray}
To compute the conserved mass of the spacetime, we choose a timelike Killing
vector field $\xi $ on the boundary surface ${\cal B}$ with the metric (\ref
{MetB}). Then the quasilocal conserved mass can be written as
\begin{equation}
{\cal M}=\frac{1}{8\pi }\int_{{\cal B}}d^{2}\varphi \sqrt{\sigma }\left\{
\left( K_{ab}-Kh_{ab}\right) -\left( K_{ab}^{0}-K^{0}h_{ab}^{0}\right)
\right\} n^{a}\xi ^{b},
\end{equation}
where $\sigma $ is the determinant of the metric (\ref{MetB}), $K_{ab}^{0}$
is the extrinsic curvature of the background metric and $n^{a}$ is the
timelike unit normal vector to the boundary ${\cal B}$. In the context
of counterterm method, the limit in which the boundary ${\cal B}$ becomes
infinite (${\cal B}_{\infty }$) is taken, and the counterterm prescription
ensures that the action and conserved charges are finite. Although the
function $r(\cal R)$ cannot be obtained explicitly, but at large $R$ this can be
done. Thus, one can calculate the mass through the use of the
above modified Brown and York formalism as
\begin{equation}
M=\frac{(1+\alpha ^{2})^{2}e^{2}+(1-\alpha ^{2})r_{0}^{2}%
}{2(1+\alpha ^{2})r_{0}}. \label{Mass}
\end{equation}
In order to check the correctness of the above mass, we consider the
first law of thermodynamics. The entropy of the dilaton
black hole typically satisfies the so called area law of the entropy \cite{Beck}.
This near universal law which is applied to almost all kinds of
black holes, including dilaton black holes, in Einstein gravity \cite{hunt} gives
\begin{equation}
S=\pi r_+^2\left( 1-\frac{r_{0}}{r_+}\right) ^{2\alpha ^{2}/(1+\alpha ^{2})}.
\end{equation}
The parameter $r_0$ may be written in term of $S$ and $r_+$ as
\begin{equation}
r_0=r_+\left\{1-\left(\frac{S}{\pi r_+^2}\right)^{(1+\alpha^2)/2\alpha^2}\right\}. \label{r0}
\end{equation}
Now using Eq. (\ref{r0}) and the fact that $f(r_+)=0$, one may find the relation between
$S$ and $r_+$ as:
\begin{equation}
\left[1-\left(\frac{S}{\pi r_+^2}\right)^{(1+\alpha^2)/2\alpha^2}\right]\left[k-\frac{r_+^2}{l^2}\left(\frac{S}{\pi r_+^2}\right)^{(3\alpha^2-1)/2\alpha^2}\right]-\frac{(1+\alpha^2)e^2}{r_+^2}=0. \label{Sr}
\end{equation}
Since $r_0$ depends on $S$ and $r_+$, the mass $M$ given in Eq. (\ref{Mass}) depends on $e$, $S$ and $r_+$.
But, $S$ and $r_+$ are related by Eq. (\ref{Sr}), and therefore $M$ may be regarded as a function of the extensive parameters
$S$ and $e$. Using the fact that $dM(S,e)=(\partial/ M\partial S)_e dS+(\partial/ M \partial e)_S de$
one may show that
\begin{equation}
dM=T_+ dS+U de,
\end{equation}
where $T_+$ is given in Eq. (\ref{T2}) and $U=e/r_+$.
Thus, the mass calculated through the modified version of Brown and York formalism satisfies the first law of thermodynamics.
It is worth to mention that the coupling constant $e$ is the same as the charge $Q$ for the solution
of Einstein Maxwell gravity.
\section{Closing Remarks}
We introduced the black hole solutions of
Einstein-Yang-Mills-dilaton theory with spherical and hyperbolic horizons and investigate their
properties. First, we presented an asymptotically flat solution in the
absence of dilaton potential, for which the horizon should be a 2-sphere.
It is worth to compare the
distinguishing features of this solution with the asymptotically flat Reissner-Nordstrom solution.
The singularity of asymptotically flat solution of Einstein-Maxwell theory
is timelike and avoidable, while the singularity of the asymptotically flat
solution presented here is null and unavoidable. This is shown in the
Penrose diagram of the solution. Also, we found that one cannot have an
extreme asymptotically flat black hole, while the extreme Reissner-Nordstrom black hole exists.
Second, we investigated the asymptotically non-flat solutions in the presence of
a dilaton potential. In this case, the horizon can also be a hyperbola,
which is known as the topological solution. The asymptotic behavior
of these solutions with spherical and hyperbolic horizons are neither
flat nor AdS. The singularities of
these solutions are null, but they are
timelike and therefore avoidable. These solutions may present a
naked singularity, an extreme black hole or a black hole with two horizons.

As we mentioned, the solutions in the presence of dilaton
potential are neither asymptotically flat nor AdS and therefore one cannot use
the Brown and York formalism or AdS/CFT counterterm method to compute the mass of the black holes.
To compute the mass of the solutions, we introduced a modified version of the
Brown and York formalism and calculated the mass.
We checked the correctness of the mass computed through
the use of the modified version of Brown and York by
investigating the first law of thermodynamics. We found that
the calculated mass satisfies the first law of thermodynamics. In this paper, we only introduced
the 4-dimensional solutions of Einstein-Yang-Mills theory in the presence
of dilaton. Since higher-dimensional black holes have attracted a lot of interest,
it is worth to introduce the higher-dimensional
solutions in EYMD theory and investigate their properties.

{\bf Acknowledgements}

This work has been supported by Research Institute for Astrophysics and
Astronomy of Maragha.


\begin{references}
\bibitem{Wit1}  M. B. Green, J. H. Schwarz, and E. Witten, Superstring
Theory (Cambridge University Press, Cambridge, England, 1987).

\bibitem{CD}  G. W. Gibbons, K. Maeda, Nucl. Phys. \textbf{B298} (1988) 741;\\
D. Brill, G. Horowitz, Phys. Lett. {\bf B262} (1991) 437;\\
D. Garfinkle, G. Horowitz, A. Strominger, Phys. Rev. D {\bf 43} (1991) 3140;\\
R. Gregory, J. Harvey, Phys. Rev. D {\bf 47} (1993) 2411.

\bibitem{AdSdil}  K. C. K. Chan, J. H. Horne, and R. B. Mann, Nucl. Phys.
\textbf{B447} (1995) 441.

\bibitem{Gao}  C. J. Gao and S. N. Zhang, Phys. Rev. D \textbf{70} (2004) 124019;
Phys. Lett. \textbf{B605} (2005) 185.

\bibitem{BK}  R. Bartnik and J. McKinnon, Phys. Rev. Lett. {\bf 61} (1988) 141.

\bibitem{YMblack}  M. S. Volkov and D. V.Galtsov, JETP Lett. {\bf 50} (1989)
346350;\\
H. P. Kunzle and A. K. M. Masood ul Alam, J. Math. Phys. {\bf %
31} (1990) 928935;\\ P. Bizon, Phys. Rev. Lett. {\bf 64} (1990) 2844.

\bibitem{Yas}  P. B. Yasskin, Phys. Rev. D {\bf 12} (1975) 2212.

\bibitem{Hal1}  S. H. Mazharimousavi and M. Halilsoy, Phys. Lett. {\bf B659}
(2008) 471.

\bibitem{BD}  N. Bostani, and M. H. Dehghani, arXiv:0908.0661, Mod. Phys. Lett. A, to appear;\\
M. H. Dehghani, N. Bostani and R. Pourhasan, arXiv:0908.0663, Int. J. Mod. Phys. D, to appear.

\bibitem{Volkov}  M. Volkov and D. V. Gal'tsov Phys. Rept. {\bf 319} (1999) 1.

\bibitem{Radu1}  E. Radu and D. H. Tchrakian, Gravitating Yang--Mills fields in all dimensions, arXiv:0907.1452 [gr-qc].

\bibitem{Vol2}  T. Torii, K. I. Maeda and T. Tachizawa, Phys. Rev. D {\bf 52} (1995) R4272;\\
M. S. Volkov et. al., Phys. Rev. D. {\bf 54} (1996) 7243;\\ R. B. Mann, E. Radu and D.
H. Tchrakian, Phys. Rev. D. {\bf 74} (2006) 064015.

\bibitem{Eliza}  J. Bjoraker and Y. Hosotani, Phys. Rev. Lett. {\bf 84} (2000)
1853;\\ J. E. Baxter and E. Winstanley, Class. Quant. Grav. {\bf 25} (2008)
245014.

\bibitem{EYMd}  G. Lavrelashvili and D. Maison, Nucl. Phys. \textbf{B410} (1993)
407;\\ E. E. Donets and D. V. Galtsov, Phys. Lett. B 302 (1993) 411;\\ P. Bizon,
Acta Phys. Polonica B 24 (1993) 1209;\\ T. Torii and K. Maeda, Phys. Rev. D 48
(1993) 1643;\\ Y. Brihaye and E. Radu, Phys. Lett. \textbf{B636}, 212 (2006);\\
E. Radu, Y. Shnir and D. H. Tchrakian, Phys. Rev. D \textbf{75} (2007) 045003;\\
M. Cvetic, H. Lu and C. N. Pope, arXiv:09080131.

\bibitem{Cai}  R. G. Cai, J.Y. Ji, and K. S. Soh, Phys. Rev. D 57
(1998) 6547;\\ R. G. Cai and Y. Z. Zhang, Phys. Rev. D. 64 (2001) 104015;\\ M. H.
Dehghani and N. Farhangkhah, Phys. Rev. D. \textbf{71} (2005) 064008;\\ M. H. Dehghani,
Phys. Rev. D. \textbf{71} (2005) 064010;\\ A. Sheykhi, M. Allahverdizadeh, Phys. Rev. D {\bf 78},
064073 (2008).

\bibitem{Wu}  T. T. Wu and C. N. Yang, {\it Properties of Matter Under
Unusual Conditions}, edited by H. Mark and S. Fenbach (Interscience, New
York, 1969), p. 349.

\bibitem{Gid}  S. B. Giddings, Phys. Rev. D {\bf 68} (2003) 026006;\\
E. Radu, D. H. Tchrakian, Class. Quantum Grav. {\bf 22} (2005) 879.

\bibitem{BY}  J. Brown and J. York, Phys. Rev. D {\bf 47} (1993) 1407;\\
J. D. Brown, J. Creighton, and R. B. Mann, Phys. Rev. D {\bf 50} (1994) 6394.

\bibitem{Beck}  J. D. Beckenstein, Phys. Rev. D {\bf 7}, 2333 (1973);\newline
S. W. Hawking, Nature (London) {\bf 248}, 30 (1974);\newline
G. W. Gibbons and S. W. Hawking, Phys. Rev. D {\bf 15}, 2738 (1977).

\bibitem{hunt}  C. J. Hunter, Phys. Rev. D {\bf 59} (1999) 024009;\newline
S. W. Hawking, C. J. Hunter and D. N. Page, Phys. Rev. D {\bf 59} (1999) 044033;\newline
R. B. Mann Phys. Rev. D {\bf 60} (1999) 104047.
\end{references}
\end{document}